\documentclass[fleqn,usenatbib]{mnras}

\usepackage{newtxtext,newtxmath}

\usepackage[T1]{fontenc}
\usepackage{ae,aecompl}

\usepackage{pdflscape} 

\usepackage{graphicx} 
\usepackage{amsmath}  
\usepackage{amssymb}  
\usepackage{threeparttable} 

\title[Fundamental Effective Temperature of AI Phoenicis]{Fundamental effective temperature measurements for eclipsing binary stars \\ I. Development of the method and application to AI Phoenicis}

\author[N. J. Miller et al.]{
N. J. Miller,$^{1}$\thanks{E-mail: n.j.miller1@keele.ac.uk}
P. F. L. Maxted$^{1}$\thanks{E-mail: p.maxted@keele.ac.uk}
and B. Smalley$^{1}$
\\
$^{1}$Astrophysics Group, Keele University, Keele, Staffordshire, ST5 5BG, UK\\
}

\date{Accepted 2020 July 16. Received 2020 June 30; in original form 2020 March 27}

\pubyear{2020}

\begin{document}
\label{firstpage}
\pagerange{\pageref{firstpage}--\pageref{lastpage}}
\maketitle

\begin{abstract}
Stars with accurate and precise effective temperature (T$_{\rm eff}$) measurements are needed to test stellar atmosphere models and calibrate empirical methods to determine T$_{\rm eff}$. There are few standard stars currently available to calibrate temperature indicators for dwarf stars. Gaia parallaxes now make it possible, in principle, to measure T$_{\rm eff}$ for many dwarf stars in eclipsing binaries.
We aim to develop a method that uses high-precision measurements of detached eclipsing binary stars, Gaia parallaxes and multi-wavelength photometry to obtain accurate and precise fundamental effective temperatures that can be used to establish a set of benchmark stars. We select the well-studied binary AI Phoenicis to test our method, since it has very precise absolute parameters and extensive archival photometry.
The method uses the stellar radii and parallax for stars in eclipsing binaries. We use a Bayesian approach to obtain the integrated bolometric fluxes for the two stars from observed magnitudes, colours and flux ratios.
The fundamental effective temperature of two stars in AI Phoenicis are $6199\pm22$\,K for the F7\,V component and $5094\pm16$\,K for the K0\,IV component. The zero-point error in the flux scale leads to a systematic error of only 0.2\% ($\approx$ 11\,K) in T$_{\rm eff}$. We find that these results are robust against the details of the analysis, such as the choice of model spectra.
Our method can be applied to eclipsing binary stars with radius, parallax and photometric measurements across a range of wavelengths. Stars with fundamental effective temperatures determined with this method can be used as benchmarks in future surveys.

\end{abstract}

\begin{keywords}
stars: solar-type -- binaries: eclipsing -- stars: fundamental parameters
\end{keywords}



\section{Introduction}

Detached eclipsing binary stars (DEBS) that are also double-lined spectroscopic binaries (SB2s) are ideal benchmark stars because the mass and radius of both stars can be directly measured to high precision and accuracy without a strong dependence on models. 
With high-quality light curves from space-based instrumentation such as Kepler and TESS combined with radial velocity measurements from echelle spectrographs, many DEBS now have radii and masses measured to an accuracy of 1\% or better \citep{2015ASPC..496..164S}. Some systems have masses and radii of both components measured to better than 0.5\% \citep[e.g.][]{2019MNRAS.484..451H}.
The study of solar (late) type DEBS is of particular importance. In the Geneva-Copenhagen Survey III, F/G dwarfs and K subgiants are the most suitable stars for estimating age \citep{2009A&A...501..941H}. FGK stars are the most numerous type observed by Gaia, with 86\% of all stars V<20 expected to be FGK stars \citep{2005ESASP.576...83R}. Most exoplanets to date have been found around FGK stars (and M-dwarfs) \citep{2013PNAS..11019273P,2013ApJ...766...81F}.
Asteroseismology of planet host stars using Kepler \citep[e.g.][]{2012Sci...337..556C} and the upcoming PLATO mission \citep{2014ExA....38..249R} is most successful on F/G dwarfs and G/K subgiants.

While the best stellar mass and radius measurements are now very accurate, effective temperature measurements remain significant sources of uncertainty. \citet{2019ARA&A..57..571J} found that most FGK type stars (excluding the Sun) have calculated effective temperatures accurate to $200-300$\,K, with none more accurate than 50\,K. 
Errors in T$_{\rm eff}$ are now the dominant source of uncertainty in calibrating stellar models \citep{2016A&A...587A..16V}. 
Furthermore, \citet{2018A&A...620A.168V} found that introducing a systematic T$_{\rm eff}$ error of $\pm 150$\,K has a significant effect on the uncertainty of reconstructed asteroseismic ages.
Most effective temperature measurements use a temperature scale defined by a set of benchmark stars \citep[e.g.][]{2015A&A...582A..49H}. 
Data-driven approaches and machine learning methods are increasingly being used for the analysis of large data sets from massive spectroscopic surveys \citep{2018A&A...616A...8A, 2015ApJ...808...16N}. These are trained and calibrated on data with classical determinations of parameters. There is no physics in these data-driven methods, so they rely heavily on benchmark stars to establish how features in the data relate to physical quantities, such as T$_{\rm eff}$.
There are several considerations when making a reliable T$_{\rm eff}$ scale. 
First, the accuracy. Typically, for calibrators (FGK dwarf benchmarks) the accuracy is 100\,K. There are few stars with T$_{\rm eff}$ measurements more accurate than this (e.g. Sun). 
Secondly, consistency is needed between surveys, methods, and stars of different [Fe/H], [$\alpha$/Fe], and logg. 
Finally, precise T$_{\rm eff}$ measurements is necessary, but the current state of measurements suggests that precision is not such an urgent issue to address. 

Most effective temperature estimates are made indirectly, using spectroscopic or photometric temperature indicators.
In general, spectroscopic determinations of effective temperature can give precise values with typical uncertainties of 50\,K \citet{2019ARA&A..57..571J}, but these depend considerably on stellar atmosphere models.
As an alternative to using high quality spectroscopy, effective temperatures are often determined using colour-temperature relations based on the infrared flux method (IRFM) \citep[e.g.][]{2010A&A...512A..54C}. The IRFM uses widely available photometry to obtain a relatively accurate T$_{\rm eff}$ mostly free from model dependence. However, results from this method suffer from uncertainties in interstellar extinction, flux calibrations and stars with anomalous abundances --- the variation in T$_{\rm eff}$ values for the same star with different photometry, extinction law or colour-temperature relation is typically 100\,K \citep{2011A&A...530A.138C,2019ARA&A..57..571J}.

The most accurate, model-independent determinations of effective temperature come from a fundamental approach based on the Stefan-Boltzmann law. This approach requires measurements of the angular diameter and absolute flux. However, there are few stars for which the necessary data exist and are accurate. 
\citet{2015A&A...582A..49H} used interferometric angular diameters along with bolometric fluxes from integrated observed spectral energy distributions to calculate T$_{\rm eff}$ values for a sample of FGK stars to a precision of 1.5\% or better.
The accuracy of their values of $\theta$ and F$_{\rm bol}$ were up to 3\% and 5\% respectively, corresponding to errors in T$_{\rm eff}$ of 1.5\% and 1\%, i.e. approximately 100\,K for a solar-type star.
This approach is only possible for nearby stars with suitably large angular diameters.
Angular diameter can also be inferred from radius measured in an eclipsing binary, along with a parallax. This technique has not been applied much to date because good parallax measurements have not been available for many EBs. For example, \citet{1998A&A...330..600R} used a sample of well-studied DEBS and Hipparcos parallax measurements to determine T$_{\rm eff}$ to $1-10$\%. The results from this study suffered from uncertainties dominated by Hipparcos parallax and bolometric corrections, which were used along with visual apparent magnitudes to obtain bolometric flux. 

The release of Gaia DR2 has provided precise parallax measurements of 1.3 billion stars \citep{2018A&A...616A...1G}. This presents an opportunity to revisit a fundamental approach for determining the effective temperatures of a large sample of DEBS, and in future work, the possibility of constructing a sample of FGK benchmark stars with accurate and precise mass, radius and effective temperature measurements.
In this paper we will describe a method for calculating the fundamental T$_{\rm eff}$ of DEBS using angular diameters derived from stellar radii and parallax, and bolometric flux calculated by distorting model SEDs to fit a set of apparent magnitudes, photometric colours and flux ratios covering a wavelength range from far ultraviolet to the infrared. 
We apply this method to the well-studied DEB AI Phoenicis, an ideal case to show its potential. It is possible to get very high precision in T$_{\rm eff}$ for this system due to the very accurate radii \citep{2020arXiv200309295M}, precise parallax measurements from two independent sources, good constraints on UV flux from IUE, and a tight upper limit on reddening. Based on these results, we discuss the prospects for this method in terms of testing stellar models and creating benchmark stars.

\section{Method}
Our method is based on the definition of the effective temperature for a star of radius $R$ and  luminosity $L$, viz. 
$$L=4\pi R^2 \sigma_{\rm SB} {\rm T}_{\rm eff}^4,$$
where $\sigma_{\rm SB}$ is the Stefan-Boltzmann constant. For a binary star at distance $d$, i.e. with parallax $\varpi=1/d$, the flux corrected for extinction observed at the top of Earth's atmosphere is
$$f_{0,b}= f_{0,1}+f_{0,2}=\frac{\sigma_{\rm SB}}{4}\left[\theta_1^2{\rm T}_{\rm eff,1}^4 + \theta_2^2{\rm T}_{\rm eff,2}^4\right],$$
where $\theta_1=2R_1\varpi$ is the angular diameter of star 1, and similarly for star 2.
The radius here refers to the the Rosseland radius, which is not necessarily identical to the radius obtained from the analysis of the light curve. Any difference between these definitions of the radius will be on the order of the atmospheric scale height, so will only be a significant difference for stars with very precise radii.

We use observed apparent magnitudes in a number of ultraviolet, optical and infrared photometric systems to measure the integrated (bolometric) flux $f_{0,b}$. 
The method outlined below tries to find a balance between the contributions from the data and the physics: bolometric luminosity (over the whole spectrum) is only obtainable from band-limited photometric measurements if the spectral energy distribution (SED) is known, which requires knowledge of T$_{\rm eff}$. 
To break this circular argument we use Legendre polynomials to distort the model SED for each star and produce the functions that are integrated to predict observed magnitudes, flux ratios, etc. We use Legendre polynomials as the basis functions for this distortion because they are smooth functions that are easy to compute and that can be normalised over the wavelength range of interest. The resulting integrating function will therefore have realistic small-scale features (absorption lines, absorption edges and molecular bands) determined by the model atmosphere, but the broad shape of the function determined by the data.
By using good models SEDs and sufficient data to constrain the shape of the flux distribution, the integrating functions should be very close to the true SED of each star.
This approach is in part motivated by \citet{2015A&A...582A..49H}, in which the authors note a 4\% difference in F$_{\rm bol}$ obtained from integrating observed and model spectra for K and M dwarfs. 

The integration is done using the following function to represent the shape of the underlying SED for each star:
\begin{equation}
 \tilde{f}_{\lambda,i} = f^m_{\lambda,i}\,\times\,\Delta_i(x) =  f^m_{\lambda,i}\,\times\,\left(d_{0,i}+\sum_{j=1}^{N_{\Delta}}d_{j,i}P_{j}(x)\right).
\end{equation}
Here, $f^m_{\lambda,i}$ is the SED from a stellar model atmosphere defined over the range ($\lambda_{\rm min}, \lambda_{\rm max}$) for star $i$, $P_{j}$ is the $j^{\rm th}$ Legendre polynomial and $$x=2(\lambda-\lambda_{\rm min})/(\lambda_{\rm max}-\lambda_{\rm min})-1.$$ 
$\Delta_i(x)$ is the distortion function that is applies to the model SED for star $i$ to obtain the integrating function, $\tilde{f}_{\lambda,i}$. This distortion function is a linear combination of Legendre polynomials with coefficients $d_{j,i}$. These distortion coefficients are determined by finding the best fit to the available data with additional constraints, if necessary, to ensure the integrating functions are realistic.
$N_{\Delta}$ is the number of distortion coefficients, which we assume here is the same for both stars. Choosing the best value of $N_{\Delta}$ is a matter of trial-and-error --- this is discussed further below.
The constant $d_{0,i}$ is calculated such that $\Delta_i(x) =1$ at $\lambda = 5556$\,\AA. The tilde symbol here denotes that this function is normalised, i.e.
 $$\int_{\lambda_{\rm min}}^{\lambda_{\rm max}} \tilde{f}_{\lambda} d\lambda = 1.$$
The limits $\lambda_{\rm min} = 1000$\,\AA\ and $\lambda_{\rm max} = 30$\,$\mu$m are chosen so that at least 99.8\% of the flux from the model SED is included within these limits. 
The ``distortion coefficients'' $d_{j,i}$ are included so that the overall shape of the SED for each star is determined by the data, not the model.

Since $\tilde{f}_{\lambda,i}$ is normalised we can use the following function to represent the observed flux at the top of Earth's atmosphere from star $i$:
$$f_{\lambda,i}=\frac{\sigma_{\rm SB}}{4}\theta_i^2{\rm T}_{{\rm eff},i}^4\tilde{f}_{\lambda,i}{\rm A}_{\lambda}, $$
where A$_{\lambda}$ is the wavelength-dependent extinction due to interstellar dust. We use the extinction law from \citet{1989ApJ...345..245C} for diffuse dust with $\rm  R(\rm V) = \rm A(\rm V)/\rm E(\rm B-\rm V) = 3.1$.
These are the functions that are integrated over the appropriate response functions to calculate flux ratios, photometric colours and apparent magnitudes in each photometric system. 

We use a Bayesian approach to find the posterior distributions for T$_{\rm eff,1}$ and T$_{\rm eff,2}$. We frame this problem in terms of a set of model parameters, $M$,  and the data, $D$, so that the posterior probability distribution is
$$ P(M|D) \propto P(D|M) P(M),$$
where $P(M)$ is the prior probability distribution for $M$.  The likelihood $P(D|M)$ can be divided into four parts according to the type of the data being used:
$$P(D|M) \propto {\cal L}_{\theta}\,{\cal L}_m\,{\cal L}_x\,{\cal L}_{\ell}.$$
Since we are only interested in relative values of the overall likelihood we omit any constant normalisation factors when we calculate these individual contributions to the likelihoods.
 
The angular diameters $\theta_1\pm\sigma_{\theta, 1}$ and $\theta_2\pm\sigma_{\theta,2}$ are not independent because they are both calculated using a single value of the parallax and, in general, $R_1$ and $R_2$ will also be correlated, i.e., the joint probability distribution $P(\theta_1,\theta_2)$ has a non-zero correlation coefficient, $\rho$.  For a model where the angular diameters of the stars are $\theta^{\prime}_1$ and $\theta^{\prime}_2$, we account for this correlation by using the following expression to calculate the log-likelihood ${\cal L}_{\theta}$:
 $$\log_e {\cal L}_{\theta}=-\frac{z}{2(1-\rho^2)}, $$
 where 
 $$z =\frac{(\theta^{\prime}_1 - \theta_1)^2}{\sigma_{\theta,1}^2}
-\frac{2\rho(\theta^{\prime}_1-\theta_1)(\theta^{\prime}_2-\theta_2)}{\sigma_{\theta,1}\sigma_{\theta,2}}
+\frac{(\theta^{\prime}_2 - \theta_2)^2}{\sigma_{\theta,2}^2}.$$

For each observed apparent magnitude $m_k \pm \sigma_{m,k}$ we predict a synthetic magnitude $m^{\prime}_k$ by numerical integration of the binary SED, $f_{\lambda,b}=f_{\lambda,1}+f_{\lambda,2}$, weighted by the response function, $R_m(\lambda)$. The details of how this synthetic photometry is calculated vary between different photometric systems, but always requires a zero-point magnitude which has some uncertainty, i.e. a standard error $\sigma_{{\rm zp},k}$. In addition to this zero-point  error, there will be additional sources of error that are difficult to characterise, e.g. intrinsic stellar variability, errors in the response function, errors in the stellar models, etc. We quantify this additional noise with a single parameter $\sigma_{{\rm ext},m}$. The likelihood ${\cal L}_m$ is then given by 
$$\log_e {\cal L}_m=-0.5\times\sum_k \left(w_{k,m} (m^{\prime}_k-m_k)^2 - \log_e(w_{k,m})\right), $$
where $w_{k,m} = 1/(\sigma_{m,k}^2+\sigma_{{\rm zp},k}^2+\sigma_{{\rm ext},m}^2)$. 

For an observed photometric index $x_k\pm\sigma_{x,k}$ and a model that predicts an index value $x^{\prime}_k$, the likelihood is calculated using 
$$\log_e {\cal L}_x=-0.5\times\sum_k\left( w_{k,x}\,(x^{\prime}_k-x_k)^2 - \log_e(w_{k,x})\right), $$
where $w_k =1/(\sigma_{x,k}^2+\sigma_{{\rm ext},c}^2)$, and $\sigma_{{\rm ext},c}$ is a parameter that quantifies external error sources in photometric indices.

It is essential to have measurements of the flux ratio at a number of different wavelengths in order to calculate accurate effective temperatures for both stars independently. We assume that these flux ratios, $\ell_k\pm\sigma_{\ell,k}$, are also affected by an additional noise source with standard deviation $\sigma_{\rm ext, \ell}$, partly because of the reasons listed above for apparent magnitudes, but also because the errors in these ratios derived from the light curve analysis may be underestimated. With this assumption, the contribution to the likelihood for a model that predicts flux ratios $\ell^{\prime}_k$ is
 $$\log_e {\cal L}_{\ell}=-0.5\times\sum_k\left( w_{k,\ell}(\ell^{\prime}_k-\ell_k)^2- \log_e(w_{k,\ell})\right),$$
 where $w_{k,\ell}=1/(\sigma_{\ell,k}^2+\sigma_{{\rm ext},\ell}^2).$

The first eight free parameters in our model, $M$, are  ${\rm T}_{\rm eff,1}$, ${\rm T}_{\rm eff,2}$,  $\theta^{\prime}_1$, $\theta^{\prime}_2$,  E(B$-$V), $\sigma_{{\rm ext},m}$, $\sigma_{{\rm ext},\ell}$ and $\sigma_{{\rm ext},c}$. We find that these parameters are all well-constrained by the data so we use an improper uniform prior that requires these quantities to be positive-definite, but that has no upper bound. For the distortion coefficients we set a uniform prior over the range $[-1, 1]$. We use {\sc emcee} \citep{2013PASP..125..306F}, a {\sc python} implementation of an affine-invariant Markov chain Monte Carlo (MCMC) ensemble sampler, to calculate the posterior probability distribution (PPD) of these model parameters, i.e., to generate a large sample of points drawn from probability distribution $P(M|D)$.

\begin{table}
\caption{Data for AI~Phe used in our analysis. }
\label{DataTable}
\begin{center}
\begin{tabular}{@{}lrl}
\hline
\multicolumn{1}{@{}l}{Quantity}&
\multicolumn{1}{l}{Value}&
\multicolumn{1}{l}{Source}\\
\noalign{\smallskip}
\hline
\noalign{\smallskip}
Parallax, $\varpi$ & $5.885\pm 0.019$\,mas & See text.\\
\noalign{\smallskip}
Radius, R$_1$ & $1.8036\pm 0.0022$\,R$_{\odot}$ &\citet{2020arXiv200309295M}$^1$\\
Radius, R$_2$ & $2.9303\pm 0.0023$\,R$_{\odot}$ &~~~~"\\
\noalign{\smallskip}
\hline
\multicolumn{3}{@{}l}{Apparent magnitude} \\
~FUV  &$ 20.473\pm 0.160  $& GALEX  \\
~u320  &$ 10.734\pm 0.035$& This work \\
~u220n &$ 13.932\pm 0.100$&~~~" \\
~u220w &$ 14.066\pm 0.119$&~~~" \\
~G    &$  8.443\pm 0.0002 $& Gaia DR2 \\
~BP   &$  8.798\pm  0.001 $& ~~~" \\
~RP   &$  7.914\pm  0.001 $& ~~~" \\
~J    &$  7.301\pm  0.023 $& 2MASS  \\
~H    &$  6.935\pm  0.034 $& ~~~"  \\ 
~Ks   &$  6.819\pm  0.026 $& ~~~"  \\
~W1   &$  6.747\pm  0.037 $& WISE  \\ 
~W2   &$  6.830\pm  0.022 $& ~~~"  \\
~W3   &$  6.811\pm  0.016 $& ~~~"  \\
~W4   &$  6.768\pm  0.061 $& ~~~" \\
\hline
\multicolumn{3}{@{}l}{Flux ratios} \\
~u320  &$ 0.342\pm 0.042$& This work \\
~u220n &$ 0.030\pm 0.066$&~~~" \\
~u220w &$ 0.059\pm 0.090$&~~~" \\
~U    &$  0.442 \pm 0.021$& \protect{\citet{1988A&A...196..128A}}  \\
~U    &$  0.446 \pm 0.020$& \multicolumn{1}{c}{"}  \\
~B    &$  0.725 \pm 0.011$& \multicolumn{1}{c}{"}  \\
~B    &$  0.727 \pm 0.011$& \multicolumn{1}{c}{"}  \\
~V    &$  1.011 \pm 0.009$& \multicolumn{1}{c}{"}  \\
~V    &$  1.011 \pm 0.009$& \multicolumn{1}{c}{"}  \\
~R    &$  1.197 \pm 0.024$& \multicolumn{1}{c}{"}  \\
~R    &$  1.198 \pm 0.024$& \multicolumn{1}{c}{"}  \\
~I    &$  1.406 \pm 0.023$& \multicolumn{1}{c}{"}  \\
~I    &$  1.406 \pm 0.023$& \multicolumn{1}{c}{"}  \\
~u     &$  0.475 \pm 0.017$& \multicolumn{1}{c}{"}  \\
~v     &$  0.624 \pm 0.009$& \multicolumn{1}{c}{"}  \\
~b     &$  0.870 \pm 0.006$& \multicolumn{1}{c}{"}  \\
~y     &$  1.036 \pm 0.007$& \multicolumn{1}{c}{"}  \\
~TESS  &$  1.319 \pm 0.001$ & \citet{2020arXiv200309295M}\\
~H     &$  2.012 \pm 0.010$ & \protect{\citet{2019A&A...632A..31G}} \\
\hline
\multicolumn{3}{@{}l}{Str\"{o}mgren photometry} \\
~(b$-$y) &$0.431\pm 0.0037$  & \protect{\citet{2009A&A...501..941H}} \\
~m$_1$   &$0.209\pm 0.0041$ &~~~" \\
~c$_1$  &$0.356\pm  0.0066 $&~~~" \\
~(b$-$y) &$0.424\pm 0.0037$ & \protect{\citet{1978IBVS.1419....1R}} \\
~m$_1$   &$0.219\pm 0.0041$ &~~~" \\
~c$_1$  &$0.357\pm  0.0066 $&~~~" \\
\hline
\multicolumn{3}{@{}l}{Derived quantities} \\
$\theta_1$ & $0.0988\pm 0.0004$  & \protect{$2R_1 \varpi$} \\
$\theta_2$ & $0.1606\pm 0.0005$  & \protect{$2R_2 \varpi$} \\
\noalign{\smallskip}
\hline
\end{tabular}
\begin{tablenotes}
$^1$Including correction from apparent disc radius to Rosseland radius.
\end{tablenotes}
\end{center}
\end{table}

\begin{figure}
\centering
\includegraphics[width=0.475\textwidth]{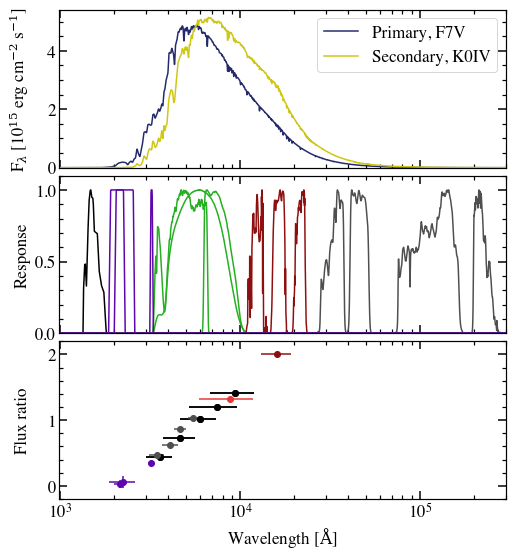}
\caption{Multi-panel plot showing data we used in our analysis. 
\textit{Top}: BT-Settl 1D model SEDs for the two stars: primary ${\rm T}_{\rm eff,1}$ = 6200\,K, logg = 3.5, [Fe/H] = -0.14, [$\alpha$/Fe] = 0.06; secondary ${\rm T}_{\rm eff,2}$ = 5100\,K, logg = 4.0, [Fe/H] = -0.14, [$\alpha$/Fe] = 0.06. These have been brought onto the same scale by multiplying by $r_1^2$ and $r_2^2$ respectively, where $r=R/a$.
\textit{Middle}: Filter response profiles for apparent magnitude data. In order of increasing wavelength: GALEX FUV, IUE (see Section \ref{Section:IUE}), Gaia BP, G and RP, 2MASS JHKs, WISE W1-4.
\textit{Bottom}: Observed flux ratio values and standard errors from IUE, Stromgren uvby, Johnson UBVRI, TESS, and H plotted over the associated bandpass, centred on pivot wavelength. }
\label{InputData}%
\end{figure}

\section{Application to AI Phe}

We selected AI~Phe as the first eclipsing binary to  analyse with our method because this is a very well studied eclipsing binary that is moderately bright (V=8.6) for which good-quality light curves in several photometric bands are available from the near-ultraviolet (NUV) to the I-band, including a very high quality light curve from the TESS mission. \citet{2020arXiv200309295M} analysed the TESS light curve of AI~Phe using several different methods and, in combination with  spectroscopic  orbits from 3 independent sources, were able to measure the masses and radii of both stars to an accuracy of better than 0.2\%.
This very high accuracy in the stellar radii is possible because AI Phe is a bright system with stars of similar brightness where the eclipses are total. This gives a direct measurement of the flux ratio for the binary from the depth of the eclipse where one star is completely occulted, and strong constraints on the geometry of the binary from the contact points of the eclipse. Limb-darkening does add some uncertainty to the measurements of the radii. Maxted~et~al., accounted for this in their analysis by modelled the light curve using several different methods to parameterise the limb darkening. \citet{2009EGUGA..11.3961H} found that the radius of the apparent solar disc is 0.33Mm larger than its Rosseland radius. Scaling this value according to the atmospheric pressure scale height we find that the Rosseland radii of the stars in AI~Phe are approximately 0.1\% smaller than the values given in \citet{2020arXiv200309295M}. All radius values used in this paper refer to the Rosseland radius including this correction.
 AI~Phe is an important system for testing stellar evolution models of single stars \citep[e.g.,][]{1988A&A...196..128A, 1997MNRAS.289..869P,  2000MNRAS.318L..55R, 2002A&A...396..551L, 2013ApJ...776...87S, 2015ApJ...812...96G, 2017A&A...608A..62H} so we felt it would be valuable to have effective temperature estimates of comparable accuracy to the masses and radii measured by  \citet{2020arXiv200309295M}.

\subsection{Data}
The input data for our analysis of AI~Phe are listed in Table~\ref{DataTable}. For the parallax we have used the average of the orbital parallax from \citet[][$5.905\pm 0.024$\,mas]{2019A&A...632A..31G}  and the value from the Gaia DR2 catalogue ($5.8336\pm 0.0262$\,mas; Gaia collaboration \citeyear{2018A&A...616A...1G}) including the zero-point correction from \citet[][$-0.031\pm 0.011$\,mas]{2019ApJ...872...85G}. For the errors on the radii we use the sum of the random and systematic errors quoted in \citet{2020arXiv200309295M}. The Gaia G, BP and RP magnitudes are also from the Gaia DR2 catalogue and include the correction from \cite{2018MNRAS.479L.102C} that is required to make these magnitudes consistent with the CALSPEC flux scale. WISE magnitudes are taken from the All-Sky Release Catalog \citep{2012yCat.2311....0C} because we find that the photometry is more reliable for bright stars in this catalogue than the ALLWISE catalogue.\footnote{\url{http://wise2.ipac.caltech.edu/docs/release/allwise/}} The standard error estimates for the Str\"{o}mgren photometric indices are taken from \citet{1994A&AS..106..257O}.

Flux ratios from the analysis of the UBVRI and uvby light curves are taken from \citet{1988A&A...196..128A}. These should be very reliable because the primary eclipse is total, i.e. the flux ratio is determined directly from the depth of the eclipse. For the flux ratio in the TESS band we take the mean value with its standard deviation from the results presented in \citet{2020arXiv200309295M} using a variety of analysis methods. We also include the flux ratio value 2.012 measured in the H-band using the VLTI interferometer by \citet{2019A&A...632A..31G}. The bandpass for this flux ratio measurement is not well defined so we assign an nominal error to this value of 0.01, and use the 2MASS H-band to calculate the flux ratio for the measurement. 

\subsubsection{Ultraviolet photometry and flux ratios}\label{Section:IUE}
The near-UV band contains a significant portion of the total flux, therefore including photometry measurements in this region is useful for constraining the shape of the SED. AI Phe has archival GALEX NUV, FUV fluxes and IUE spectra.
Recent studies into the absolute photometric calibrations of GALEX magnitudes find a non-linear offset between archive and comparison fluxes, particularly for bright stars in the NUV \citep{2014MNRAS.438.3111C, 2019MNRAS.489.5046W}.
We compared the observed and calculated NUV magnitudes for a set of FGK dwarfs with corrected archive GALEX magnitudes and CALSPEC spectra \citep{2014PASP..126..711B}. We found that the scatter was too large for us to confidently rely on the GALEX NUV magnitude for stars of comparable brightness to AI Phe so we decided to not include it in our dataset.

\citet{1981IBVS.2060....1M} obtained a series of spectra of AI Phe during primary minimum with the International Ultraviolet Explorer (IUE) satellite. We downloaded the data from the IUE NEWSIPS archive\footnote{\url{https://archive.stsci.edu/iue/newsips/newsips.html}} and applied the wavelength-dependent corrections suggested by \citet{2019AJ....157..107B} to put the flux on the CALSPEC scale.
We created three trapezoidal filters with which to integrate the IUE spectra, the properties of which are given in Table \ref{IUEBands}. Quality flags on the data restricted the range of useful wavelengths we could use, so we placed one filter at $3200$\AA\ (u320) to capture the majority of flux, and two (narrow and wide) around the $2175$\AA\ absorption feature due to interstellar extinction (u220n, u220w) to investigate whether we could use this feature to constrain the reddening.
\begin{table}
\caption{Properties of our IUE trapezoidal band passes: pivot wavelength $\lambda_{\rm pivot}$, minimum $\lambda_{\rm min}$ and maximum $\lambda_{\rm max}$ wavelengths at which the band pass is defined, and the wavelength range over which to taper, $\lambda_{\rm soft}$.}
\label{IUEBands}
\begin{center}
\begin{tabular}{@{}lrrrr}
\hline
\multicolumn{1}{@{}l}{Band} ~~& \multicolumn{1}{l}{$\lambda_{\rm pivot}$} [\AA] ~~& \multicolumn{1}{l}{$\lambda_{\rm min}$} [\AA] ~~& \multicolumn{1}{l}{$\lambda_{\rm max}$} [\AA] ~~& \multicolumn{1}{l}{$\Delta\lambda_{\rm soft}$} [\AA]~~\\
\noalign{\smallskip}
\hline
\noalign{\smallskip}
u320  ~~& 3224 ~~& 3151 ~~& 3298 ~~& 50 ~~\\
u220w ~~& 2221 ~~& 1860 ~~& 2600 ~~& 50 ~~\\
u220n ~~& 2149 ~~& 2000 ~~& 2300 ~~& 50 ~~\\
\hline
\end{tabular}
\end{center}
\end{table}

\begin{figure}
\centering
\includegraphics[width=0.475\textwidth]{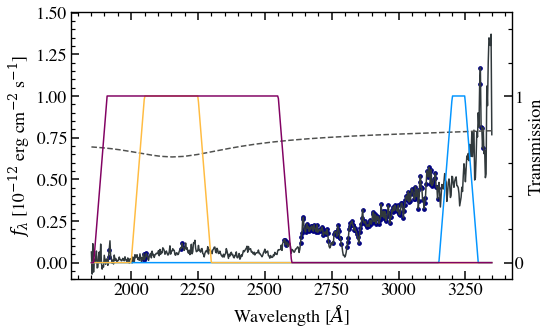}
\caption{IUE spectrum of AI Phe outside of primary minimum, with poor quality data marked (circles). Also shown are transmission profiles of the three filters u320 (purple), u220w (blue) and u220n (yellow) and an extinction profile for E(B$-$V)=0.05 (dotted line) to illustrate the location of the $2175$\AA\ absorption feature. }
\label{IUESpec}%
\end{figure}

We integrated the mean flux in each filter using Equation A5 from \cite{2012PASP..124..140B},
$$\langle{f_{\nu}}\rangle = \frac{\int{f_{\nu}(\nu) R_{\nu}(\nu) d\nu}}{\int{R_{\nu} d\nu}}, $$
and constructed light curves for each band. 
We fitted each with an {\sc ellc} light curve model \citep{2016A&A...591A.111M}, using a {\sc power2} limb darkening law with coefficients calculated from {\sc stagger} 3D atmospheric models. The primary eclipse of AI Phe is total so the choice of limb darkening model is less important, and we obtained similar results with linear and power2 limb darkening models. We used {\sc emcee} to sample the posterior distribution with the flux out of eclipse, flux ratio and log(f) as free parameters.
The maximum likelihood values for flux out of eclipse were then converted to an AB magnitude included as input data in our analysis, along with values for the flux ratio between the two stars.

\subsection{Synthetic photometry}
 Comparing synthetic magnitudes calculated from the SED of a star to observed magnitudes requires some care. We found appendix A of  \citet{2012PASP..124..140B} to be a very helpful introduction to the subject of synthetic photometry. 
 
 For the GALEX FUV band we used the response function published on the GALEX web pages.\footnote{\url{https://asd.gsfc.nasa.gov/archive/galex/tools/Resolution_Response}}. For the error on the zero-point of the GALEX AB magnitude scale we use the value 0.134\,mag from \citet{2014MNRAS.438.3111C}. For the Gaia photometry we use the revised response functions and zero-points from \citet{2018A&A...616A...4E}. The 2MASS response functions were obtained from the Explanatory Supplement to the 2MASS All Sky Data Release.\footnote{\url{https://old.ipac.caltech.edu/2mass/releases/allsky/doc/explsup.html}}. The zero-points with their standard errors are taken from \citet{2018A&A...616L...7M}. For the WISE photometry we calculate synthetic magnitudes on the AB magnitude scale and then apply the corrections to Vega magnitudes from \citet{2011ApJ...735..112J} for which the estimated error is 1.45\% (0.016 mag). For the IUE NEWSIPS spectra we adopted a zero point error of 4\% from \citet{1996AJ....111..517N}.

\subsection{Interstellar reddening}
\begin{figure}
\centering
\includegraphics[width=0.475\textwidth]{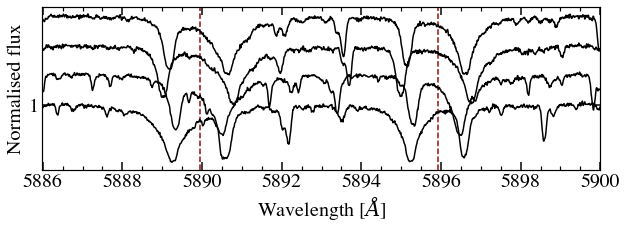}
\caption{HARPS spectra of AI~Phe in the region of the Na\,I doublet. The spectra have been normalized and vertically offset for clarity. The rest wavelengths of the Na\,I lines are indicated by vertical dashed lines. }
\label{NaISpec}%
\end{figure}

Fig.~\ref{NaISpec} shows a selection of HARPS spectra of AI~Phe obtained from the ESO Science Archive Facility in the region of the Na\,I doublet. There are no detectable interstellar absorption lines in this region and so the reddening towards AI~Phe must be very close E(B$-$V$)=0$ \citep{2010NewA...15..444K}. Accordingly, we set a Gaussian prior on the reddening E(B$-$V$)=0\pm0.005$ and set a lower limit E(B$-$V)$\ge 0$.

%

\subsection{Priors on infrared flux ratios}\label{NIRFluxRatioSect}
\begin{figure}
\centering
\includegraphics[width=0.475\textwidth]{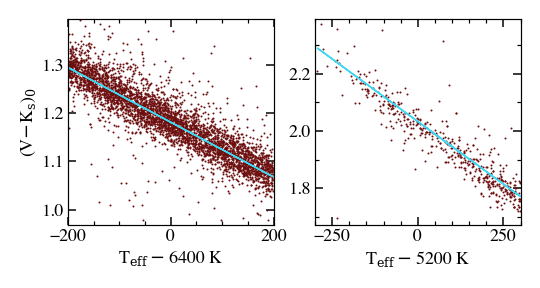}
\caption{De-reddened V$-$K colours of F-type dwarfs stars (left panel) and G/K-type sub-giants (right panel) in the solar neighbourhood. The turquoise lines show the linear fits to these data described in section~\ref{NIRFluxRatioSect}.}
\label{VKTeff}%
\end{figure}

\begin{figure}
\centering
\includegraphics[width=0.475\textwidth]{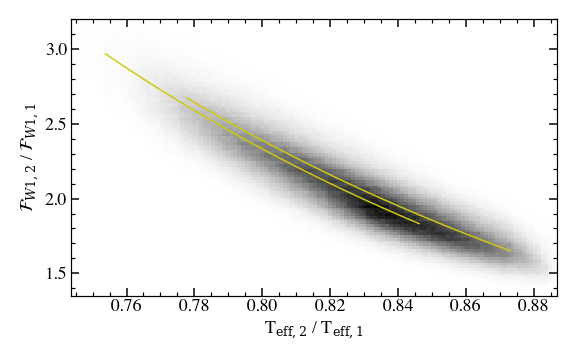}
\caption{Flux ratios in the WISE W1 band for G/K-type sub-giants relative to  F-type dwarfs stars in the solar neighbourhood. The fluxes have been scaled so that the flux ratio in the V-band is 1. The orange lines show the flux ratio predicted by our linear T$_{\rm eff}$ -- (V$-$W1) relations assuming T$_{\rm eff}$ 6300\,K or 6500\,K. }
\label{W1ratio}%
\end{figure}

The near-infrared (NIR) flux from solar-type stars compared to the total flux or the flux at optical wavelengths shows a well-defined relationship with T$_{\rm eff}$ that is almost linear and that has little dependence on metallicity or surface gravity. This is the basis of the infrared flux method \citep{1979MNRAS.188..847B, 2010A&A...512A..54C} or the use of colours such as ($\rm V-\rm K_{\rm s}$) to  estimate effective temperatures for FGK-type stars \citep{2013ApJ...771...40B}.  There is only one direct measurements of the flux ratio for AI~Phe at wavelengths longer than 1\,$\mu$m. 
We were concerned that not imposing any additional constraints on the flux ratio at these wavelengths could mean that the resulting models become unrealistic, e.g., the ($\rm V-\rm K_{\rm s}$) colours computed using samples from $P(M|D)$ might show a much large scatter than is observed in real stars. This could happen if the use of distortion coefficients allows for models where the flux from one star is unrealistically high at NIR wavelengths while the other is too low, unless some constraint in placed on the flux ratio at these wavelengths.  
We address this concern by making the assumption that the stars in AI~Phe behave similarly to other dwarf and sub-giant FGK-type stars in the solar neighbourhood in order to put some constraint on the flux ratio in the 2MASS J, H and K$_{\rm s}$ bands, and the WISE W1, W2, W3 and W4 bands. We use stars from the Geneva-Copenhagen survey \citep{2009A&A...501..941H} to define relationships between  T$_{\rm eff}$ and (V$-$J), (V$-$H), etc. The values of  T$_{\rm eff}$, E(B$-$V), $\log g$ and [Fe/H] for each star are taken from \citet{2011A&A...530A.138C}. We define separate relations for the F7\,V star and the K0\,IV star based on a different sub-sample of stars with similar properties to each. For the F7\,V star the sub-sample is defined by the following limits.
\begin{itemize}
 \item{[Fe/H] $ > -0.5$}
 \item{E(B$-$V) $ < 0.05$}
 \item{$3.5 < \log g < 4.5$}
 \item{$6200\,\rm K < \rm T_{\rm eff} < 6600\,\rm K$}
\end{itemize}
For the K0\,IV star the limits on [Fe/H] and E(B$-$V) are the same but for effective temperature and surface gravity we use the following limits.
\begin{itemize}
 \item{$3.0 < \log g < 4.5$}
 \item{$4900\,\rm K < \rm T_{\rm eff} < 5500\,\rm K$}
\end{itemize}
Both sub-samples were cross-matched with the WISE All Sky Data release \citep{2012yCat.2311....0C} using VO tools within TOPCAT \citep{2017arXiv170702160T} and matching radius of 6\,arcsec. 
Duplicate sources were removed from the sub-samples, leaving 4123 stars in the sub-sample for the F7\,V star and 556 stars in the sub-sample for the K0\,IV star. Linear relations for the  colours of these stars corrected for extinction of the form 
$$({\rm V} - X)_0 = c + m\times({\rm T}_{\rm eff} -\rm T_{\rm ref})/1000\,\rm K$$ 
were established using the median value of the sample and the robust Theil-Sen estimator of the slope, as implemented in python function {\tt scipy.stats.mstats.theilslopes}. The scatter around these relations was measured using the root-mean-square of the residuals within 5 times the median absolute deviation from the fit. The results are given in Table~\ref{nircol} and the fit to the data for (V$-$K$_{\rm s}$) is shown in Fig.~\ref{VKTeff}. The flux ratio in each band is calculated anew from the values of T$_{\rm eff,1}$, T$_{\rm eff,2}$ and the V-band flux ratio for each trail chain step, but otherwise are included in the calculation of the likelihood in the same way as the directly measured flux ratios. The flux ratio in the W1 band as a function of the effective temperature ratio is shown in Fig.~\ref{W1ratio} for every pairing of stars from the two sub-samples excluding stars more than 5-sigma from the appropriate linear T$_{\rm eff}$ -- (V$-$W1) relation. There is a well-defined correlation between the flux ratio and the effective temperature ratio relation which is accurately predicted by our linear relations, despite the fact that we have paired stars with disparate [Fe/H] and $\log g$ values.

\begin{table}
\caption{Results for robust linear fits to extinction-corrected $({\rm V} - X)_0$ colours for selected stars in the solar neighbourhood. }
\label{nircol}
\begin{center}
\begin{tabular}{@{}lrrrcrrr}
\hline
&\multicolumn{3}{c}{$\rm T_{\rm ref} = 6400$\,K }
&~&\multicolumn{3}{c}{$\rm T_{\rm ref} = 5200$\,K}\\
$X$& 
\multicolumn{1}{c}{$c$} &\multicolumn{1}{l}{$m$} & \multicolumn{1}{l}{rms}&&
\multicolumn{1}{c}{$c$} &\multicolumn{1}{l}{$m$} & \multicolumn{1}{l}{rms}\\
\noalign{\smallskip}
\hline
\noalign{\smallskip}
J  & 0.919 &$ -0.408$ & 0.015 &~& 1.511 &$ -0.605$ & 0.018 \\
H  & 1.118 &$ -0.549$ & 0.019 &~& 1.918 &$ -0.821$ & 0.027 \\
K$_{\rm s}$ & 1.181 &$ -0.564$ & 0.017 &~& 2.033 &$ -0.872$ & 0.025 \\
W1 & 1.230 &$ -0.568$ & 0.027 &~& 2.094 &$ -0.865$ & 0.035 \\
W2 & 1.234 &$ -0.547$ & 0.039 &~& 2.101 &$ -0.928$ & 0.062 \\
W3 & 1.182 &$ -0.554$ & 0.021 &~& 2.062 &$ -0.907$ & 0.036 \\
W4 & 1.225 &$ -0.519$ & 0.050 &~& 2.095 &$ -0.951$ & 0.060 \\
\noalign{\smallskip}
\hline
\end{tabular}
\end{center}
\end{table}
%

%

\begin{table}
\caption{Predicted data values and residuals for the best-fit model from Run A. The predicted apparent magnitudes are quoted together with the error on the zero-point.}
\label{Residuals}
\begin{center}
\begin{tabular}{@{}lrr}
\hline
\multicolumn{1}{@{}l}{Parameter} & \multicolumn{1}{l}{Value} & \multicolumn{1}{l}{Residual} \\
\noalign{\smallskip}
\hline
\multicolumn{3}{@{}l}{Apparent magnitude} \\ 
FUV      &$20.27\pm 0.13$&$ +0.20\pm 0.21$\\
u320     &$10.750\pm 0.043$&$ -0.016\pm 0.056$\\
u220w    &$14.029\pm 0.043$&$ +0.037\pm0.126$\\
u220n    &$14.000\pm 0.043$&$ -0.068\pm 0.109$\\
G        &$8.440\pm 0.001$&$ +0.003\pm 0.001$\\
BP       &$8.809\pm0.001$&$ -0.011\pm 0.001$\\
RP       &$7.904\pm 0.004$&$ +0.009\pm 0.004$\\
J       &$7.293\pm 0.005$&$ +0.008\pm 0.024$\\
H       &$6.910\pm 0.005$&$ +0.025\pm 0.034$\\
Ks      &$6.794\pm 0.005$&$ +0.025\pm 0.026$\\
W1      &$6.722\pm 0.002$&$ +0.025\pm 0.037$\\
W2      &$6.837\pm 0.002$&$ -0.007\pm 0.022$\\
W3     &$6.805\pm 0.002$&$ +0.006\pm 0.016$\\
W4     &$6.709\pm 0.002$&$ +0.059\pm 0.061$\\
\hline
\multicolumn{3}{@{}l}{Observed flux ratios} \\ 
u320 & 0.414 &$ 0.342 \pm 0.042$\\
u220w&  0.048&$  0.059 \pm 0.090$\\
u220n&  0.025&$  0.030 \pm 0.066$\\
U    &0.467  &$0.442 \pm0.021$\\
U    &0.467  &$0.446 \pm 0.020$\\
B    &0.731  &$0.725 \pm 0.011$\\
B    &0.731  &$0.727 \pm 0.011$\\
V    &1.005  &$1.011 \pm 0.009$\\
V    &1.005  &$1.011 \pm 0.009$\\
R    &1.206  &$1.197 \pm 0.024$\\
R    &1.206  &$1.198 \pm 0.024$\\
I    &1.374  &$1.406 \pm 0.023$\\
I    &1.374  &$1.406 \pm 0.023$\\
u    & 0.443 &$ 0.475 \pm 0.017$\\
v    & 0.635 &$ 0.624 \pm 0.009$\\
b    & 0.835 &$ 0.870 \pm 0.006$\\
y    & 1.007 &$ 1.036 \pm 0.007$\\
TESS & 1.324 &$ 1.319 \pm 0.001$\\
H    & 2.017 &$ 2.012 \pm 0.010$\\
\hline
\multicolumn{3}{@{}l}{Predicted flux ratios} \\ 
J   & 1.658 & $+0.030\pm0.023 $\\
H   & 2.017 & $+0.014\pm0.033 $\\
Ks  & 2.076 & $+0.059\pm0.030 $\\
W1  & 2.103 & $+0.053\pm0.044 $\\
W2  & 2.134 & $+0.047\pm0.073 $\\
W3  & 2.198 & $+0.003\pm0.042 $\\
W4  & 2.143 & $+0.059\pm0.078 $\\
\hline
\multicolumn{3}{@{}l}{Str\"{o}mgren photometry} \\ 
~(b$-$y) &$0.461\pm 0.004 $ & $ -0.030 \pm 0.006 $\\
~m$_1$   &$0.153\pm 0.006 $ & $ +0.056 \pm 0.007 $\\
~c$_1$   &$0.461\pm 0.008 $ & $ -0.105 \pm 0.011 $\\
~(b$-$y) &$0.461\pm 0.004 $ & $ -0.037 \pm 0.006 $\\
~m$_1$   &$0.153\pm 0.006 $ & $ +0.066 \pm 0.007 $\\
~c$_1$   &$0.461\pm 0.008 $ & $ -0.104 \pm 0.011 $\\
\hline
\multicolumn{3}{@{}l}{Angular diameters} \\ 
$\theta_1$ &$0.0989\pm 0.0004 $ & $ -0.0000 \pm 0.0006 $\\
$\theta_2$ &$0.1605\pm 0.0006 $ & $ -0.0001 \pm 0.0007 $\\
\hline
\end{tabular}
\end{center}
\end{table}

\subsection{Results}
\begin{figure}
\centering
\includegraphics[width=0.475\textwidth]{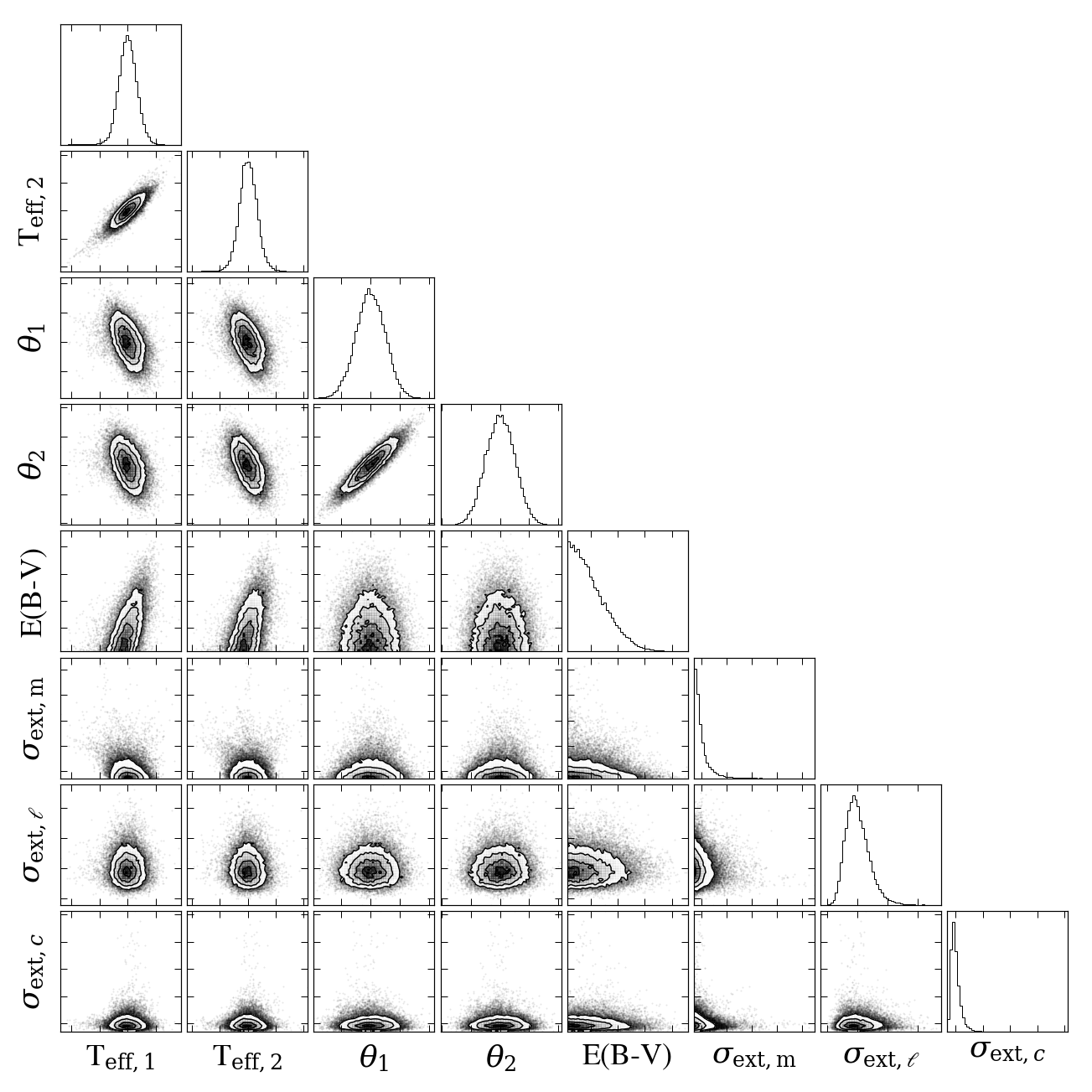}
\caption{Contour plot of the eight main parameters used in our MCMC analysis, for our primary run A in Table \ref{RunTable}.}
\label{CornerPlot}%
\end{figure}

\begin{figure}
\centering
\includegraphics[width=0.475\textwidth]{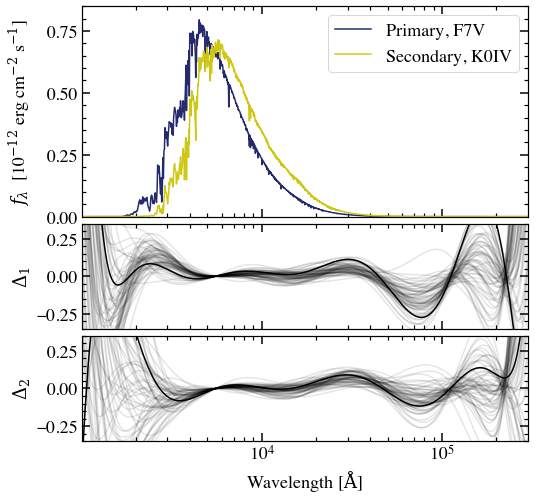}
\caption{Integrating functions and distortion functions, as defined in equation 1, for our best solution (run A). \textit{Top}: Maximum likelihood integrating functions of the two stars. \textit{Middle}: The distortion functions applied to the model SED for the primary star, showing maximum likelihood fit (thick line) and all other solutions. \textit{Lower}: Same, but for the secondary star.}
\label{OutputData}%
\end{figure}

The results of 14 fits using different sets of input parameters are given in Table \ref{RunTable}. By testing the effects of our inputs, we were able to characterise the method --- see Section \ref{sec:Discussion}. We adopt the values from run A as our final results for T$_{\rm eff}$: $6193\pm24$\,K for the F7\,V component and $5090\pm17$\,K for the K0\,IV component. We give the fundamental parameters of AI Phe from this work and \citet{2020arXiv200309295M} in Table \ref{FundamentalParams} for reference. The output integrating function and distortion functions for this solution are shown in Figure \ref{OutputData}. 
For our nominal run, we used 256 walkers randomly dispersed close to the Nelder-Mead best fit solution and ran the chain over 10000 steps with a burn-in of 4000. The resultant distributions of parameters (distortion coefficients excluded for clarity) can be seen in Figure \ref{CornerPlot}. 
%
%
%
\begin{table}
\caption{Fundamental parameters of AI Phe from this work and \citet{2020arXiv200309295M}. The radii here are Rosseland radii, calculated by applying a small correction to the radii obtained from the analysis of the eclipses by Maxted~et~al. All quantities are given in nominal solar units \citep{2016AJ....152...41P}}.
\label{FundamentalParams}
\begin{center}
\begin{tabular}{@{}lrr}
\hline
\multicolumn{1}{@{}l}{Parameter} ~~& \multicolumn{1}{l}{Value} ~~&  \multicolumn{1}{l}{Source}~~\\
\noalign{\smallskip}
\hline
\noalign{\smallskip}
M$_1$/$\mathcal{M}^{\rm N}_\odot$  ~~& $1.1938\pm0.0008$ ~~& \citet{2020arXiv200309295M} ~~\\
\noalign{\smallskip}
M$_2$/$\mathcal{M}^{\rm N}_\odot$ ~~& $1.2438\pm0.0008$ ~~& '' ~~\\
\noalign{\smallskip}
R$_1$/$\mathcal{R}^{\rm N}_\odot$ ~~& $1.8036\pm0.0022$ ~~& '' ~~\\
\noalign{\smallskip}
R$_2$/$\mathcal{R}^{\rm N}_\odot$ ~~& $2.9303\pm0.0023$ ~~& '' ~~\\
\noalign{\smallskip}
T$_1$/$\mathcal{T}^{\rm N}_\odot$ ~~& $1.074\pm0.004$ ~~& This work ~~\\
\noalign{\smallskip}
T$_2$/$\mathcal{T}^{\rm N}_\odot$ ~~& $0.882\pm0.003$ ~~& '' ~~\\
\noalign{\smallskip}
L$_1$/$\mathcal{L}^{\rm N}_\odot$ ~~& $4.329\pm0.0627$ ~~& '' ~~\\
\noalign{\smallskip}
L$_2$/$\mathcal{L}^{\rm N}_\odot$ ~~& $5.207\pm0.065$ ~~& '' ~~\\ 
\noalign{\smallskip}
\hline
\end{tabular}
\end{center}
\end{table}

Some care must be taken when using MCMC methods to explore a model parameter space with many dimensions because of the possibility that the likelihood function has more than one maximum, or that the PPD has a complex shape. Either of these possibilities makes it difficult to fully sample the PPD. This problem can be avoided by using model parameters that are closely related to features visible in the data. Legendre polynomials are orthogonal to one another, i.e., $$\int_{0}^{1} \bar{P}_{m}(x) \bar{P}_{n}(x) d x=\frac{1}{2 n+1} \delta_{m n}$$
We therefore expect that any linear combination of these polynomials that can give a good fit to the data will have coefficients with similar values.

\begin{figure}
\centering
\includegraphics[width=0.475\textwidth]{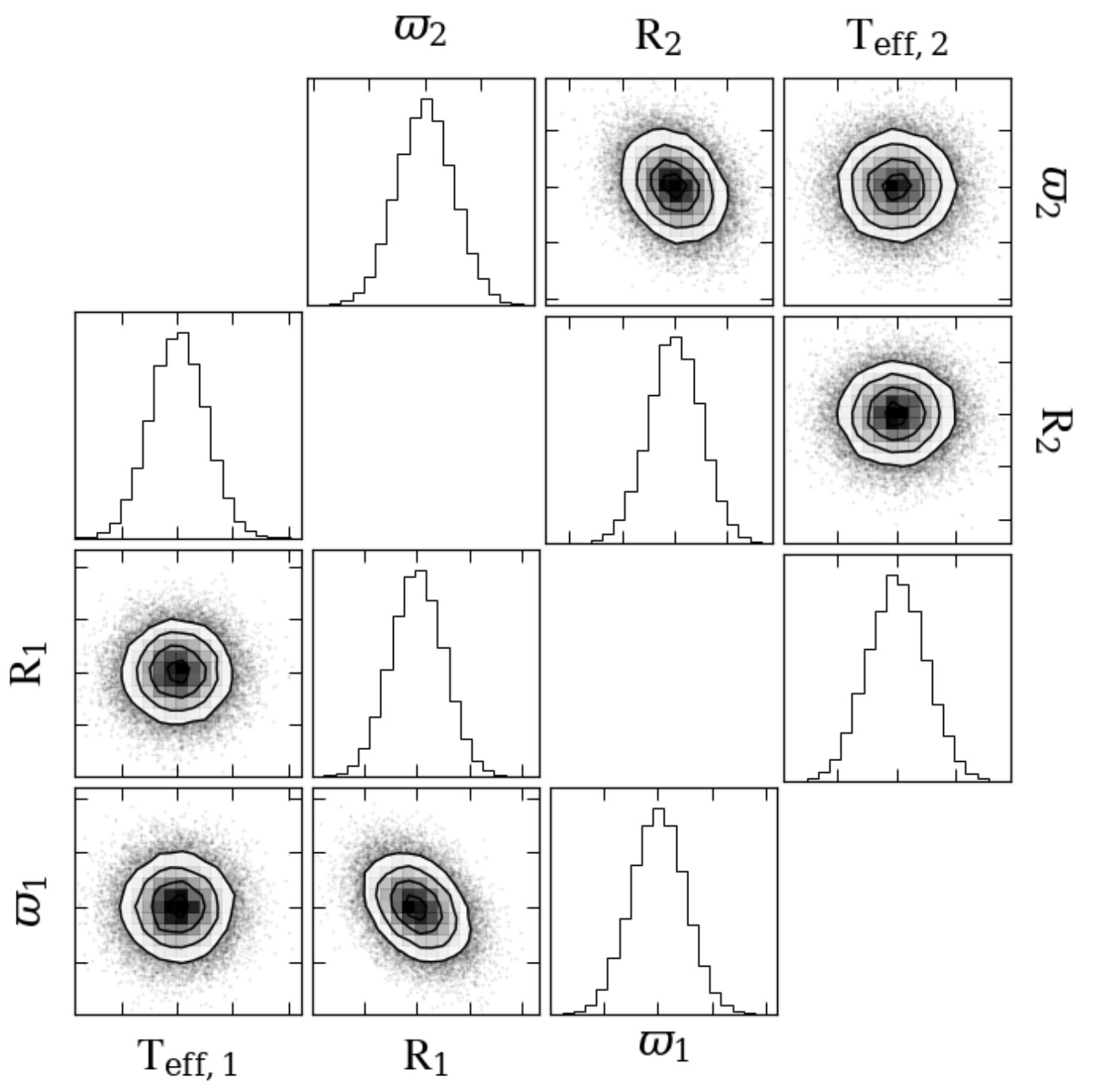}
\caption{Contour plot of the parameters used in our calculations of stellar luminosities.}
\label{LuminositiesPlot}%
\end{figure}

As expected, there is some correlation between the distortion coefficients, but the Legendre polynomials are well-behaved and the probability distributions are unimodal. Therefore the correlation is not an issue, since the affine-invariant MCMC algorithm is able to properly account for correlation in unimodal distributions.
Our effective temperatures are correlated, with a Pearson correlation coefficient of 0.805. 
An irreducible error in our analysis is the uncertainty in the flux of Vega that sets the zero-points of the photometric systems we have used. We use the value 0.5\% for the uncertainty in the flux of Vega at 5556\AA\ from \citet{2014PASP..126..711B} together with the wavelength-dependent error in the flux scale shown in their Fig.~14 to estimate the systematic error in our T$_{\rm eff}$ values from this zero-point error. By adding both these errors to the best-fit SED of each star we find that the systematic error in the integrated flux is 0.8\% for both stars, so the systematic errors in T$_{\rm eff}$ are 12\,K for the F7\,V star and 10\,K for the K0\,IV star. For most applications, the random and systematic errors can simply be added, but there are applications where the systematic error should be only added once, for example, light curve models which use the parameters T$_{\rm eff,1}$ and ${\rm T}_{\rm eff,2}$/${\rm T}_{\rm eff,1}$.
We also calculated stellar luminosities by exploring the parameter space of T$_{\rm eff}$, R and $\varpi$ for each star with {\sc emcee}. We find that $\log({\rm L_1/L_{\odot}}) = 0.636\pm0.007$ and $\log({\rm L_2/L_{\odot}}) = 0.717\pm0.005$, and correlations between the free parameters are shown in Figure \ref{LuminositiesPlot}.

\begin{landscape}

\begin{table}
\caption{Fit results from different sets of input parameters. Run A is our nominal output. Values in parentheses are 1-$\sigma$ standard errors in the final digit(s) of the preceding value. N$_{\Delta}$ is the number of distortion coefficients included per star, $\Delta \lambda$ is the size of the integrating function wavelength bins in \AA. *N.B. these parameters have a non-Gaussian probability distribution. E(B$-$V) are given as 1-$\sigma$ upper limits.
}
\label{RunTable}
\begin{threeparttable}
\begin{center}
\begin{tabular}{@{}lrrrrrrrrrrrrrrl}
\hline
\multicolumn{1}{@{}l}{Run} & 
\multicolumn{1}{@{}l}{$N_{\Delta}$} & 
\multicolumn{1}{l}{$\Delta \lambda$}&
\multicolumn{1}{l}{T$_{\rm mod,1}$}& 
\multicolumn{1}{l}{T$_{\rm mod,2}$}& 
\multicolumn{1}{l}{[Fe/H]} &
\multicolumn{1}{l}{[$\alpha$/Fe]} &
\multicolumn{1}{l}{T$_{\rm eff,1}$}& 
\multicolumn{1}{l}{T$_{\rm eff,2}$}&
\multicolumn{1}{l}{E(B$-$V)*} &
\multicolumn{1}{l}{$\sigma_{\rm ext,m}$*} &
\multicolumn{1}{l}{$\sigma_{{\rm ext,},\ell}$*} &
\multicolumn{1}{l}{$\sigma_{\rm ext,c}$*} &
\multicolumn{1}{l}{rms$_{\Delta,1}$} &
\multicolumn{1}{l}{rms$_{\Delta,2}$} &
\multicolumn{1}{l}{Notes} \\
 & & 
\multicolumn{1}{c}{[\AA]}&
\multicolumn{1}{c}{[K]}& 
\multicolumn{1}{c}{[K]}& 
\multicolumn{1}{c}{[dex]} &
\multicolumn{1}{c}{[dex]} &
\multicolumn{1}{c}{[K]}& 
\multicolumn{1}{c}{[K]}&
\multicolumn{1}{c}{[mag]} &
\multicolumn{1}{c}{[mag]} &
\multicolumn{1}{c}{[mag]} &
\multicolumn{1}{c}{[mag]} &
 & & \\
\noalign{\smallskip}
\hline
\noalign{\smallskip}
A & 10 & 20 & 6200 & 5100 & -0.14 & 0.06 & $6199\pm22$ & $5094\pm16$ & $<0.004$ & 0.01(2) & 0.02(1) & 0.09(4) & 0.036(16) & 0.035(15) & \\ 
B & 10 & 50 & 6200 & 5100 & -0.14 & 0.06 & $6197\pm22$ & $5093\pm17$ & $<0.004$ & 0.01(2) & 0.02(1) & 0.09(4) & 0.036(13) & 0.035(13) & \\ 
C & 10 & 50 & 6250 & 5050 & -0.14 & 0.06 & $6196\pm40$ & $5090\pm29$ & $<0.005$ & 0.02(4) & 0.02(1) & 0.09(3) & 0.046(43) & 0.055(37) & \\ 
D & 6 & 50 & 6200 & 5100 & -0.14 & 0.06 & $6197\pm20$ & $5095\pm15$ & $<0.004$ & 0.01(1) & 0.02(1) & 0.10(5) & 0.019(10) & 0.024(10) & \\ 
E & 14 & 50 & 6200 & 5100 & -0.14 & 0.06 & $6193\pm32$ & $5091\pm24$ & $<0.005$ & 0.03(3) & 0.02(1) & 0.08(3) & 0.061(39) & 0.058(40) & \\ 
F & 10 & 50 & 6200 & 5100 & 0 & 0 & $6192\pm23$ & $5089\pm18$ & $<0.004$ & 0.02(2) & 0.02(1) & 0.10(3) & 0.056(23) & 0.037(24) & \\ 
G & 10 & 50 & 6200 & 5100 & -0.5 & 0.2 & $6198\pm22$ & $5092\pm16$ & $<0.005$ & 0.01(2) & 0.02(1) & 0.08(3) & 0.045(12) & 0.061(12) & \\ 
H & 10 & 50 & 6200 & 5100 & -0.14 & 0.06 & $6196\pm20$ & $5091\pm15$ & $<0.004$ & 0.01(1) & 0.02(1) & 0.10(4) & 0.034(14) & 0.034(12) & \tnote{1} \\ 
I & 10 & 50 & 6200 & 5100 & -0.14 & 0.06 & $6287\pm87$ & $5146\pm56$ & $<0.03$ & 0.02(4) & 0.02(1) & 0.09(3) & 0.065(42) & 0.052(36) & \tnote{2} \\ 
J & 0 & 50 & 6200 & 5100 & -0.14 & 0.06 & $6196\pm18$ & $5098\pm13$ & $<0.005$ & 0.01(1) & 0.03(1) & 0.10(4) & -- & -- \\ 
K & 10 & 50 & 6200 & 5100 & -0.14 & 0.06 & $6332\pm120$ & $5171\pm76$ & $<0.04$ & 0.01(2) & 0.02(1) & 0.08(3) & 0.068(40) & 0.058(33) & \tnote{3} \\ 
L & 10 & 50 & 6200 & 5100 & -0.14 & 0.06 & $6194\pm23$ & $5091\pm17$ & $<0.004$ & 0.01(2) & 0.02(1) & 0.09(3) & 0.034(19) & 0.036(18) & \tnote{4} \\ 
M & 10 & 50 & 6200 & 5100 & -0.14 & 0.06 & $6217\pm86$ & $5072\pm65$ & $<0.005$ & 0.02(3) & 0.01(1) & 0.09(4) & 0.056(29) & 0.075(48) & \tnote{5} \\ 
N & 10 & 50 & 6200 & 5100 & -0.14 & 0.06 & $6196\pm21$ & $5092\pm15$ & $<0.005$ & 0.01(2) & 0.02(1) & 0.09(4) & 0.036(18) & 0.036(17) & \tnote{6} \\ 
\hline
\end{tabular}
\begin{tablenotes}
   \item[1] No NIR prior used. $^2$ No E(B$-$V) prior used. $^3$ No E(B$-$V) prior or u220n data used. $^4$ No u220n data used. $^5$ Only TESS band flux ratio used. $^6$ NIR prior model temperatures shifted up by 100\,K.
\end{tablenotes}
\end{center}
\end{threeparttable}
\end{table}
\end{landscape}

\section{Discussion}\label{sec:Discussion}

\subsection{Distortion functions}
We need to include enough distortion to avoid just performing an SED fit (see run J in Table \ref{RunTable}; uncertainties in parameters are underestimated), but not so much that the distortion of SEDs becomes non-physical.
Therefore in order to better understand how much distortion is needed in our approach, we looked into the difference between one-dimensional and three-dimensional stellar models.
Comparing BT-Settl and {\sc stagger}-3D models of similar temperatures, we found that the rms difference between the two models for the primary was 0.027, and the secondary was 0.025 --- i.e. the difference between 1-D and 3-D models is about 3\%. Most of this difference lies in the UV and, if smoothed, looks like a moderately high order (6-12) polynomial in logarithmic space. Therefore we chose to use this type of distortion in our method.

In order to be sure that we used a reasonable number of distortion coefficients, we performed some extra tests on the code to characterise the effects of distortion. We looked into the range of $0\leq {\rm N}_{\Delta}\leq 20$; being sure that on the scale of absorption lines and features, the highest order of these Legendre polynomials are linear and have no unrealistic effects.
We quantified the amount of distortion with the following:
$$ \mathrm{rms} = \sqrt{\frac{\int{F_1^m(\lambda) \Delta_1^2}{d\lambda}} {\int{F_1^m(\lambda)}{d\lambda}}} $$
\begin{figure}
\centering
\includegraphics[width=0.475\textwidth]{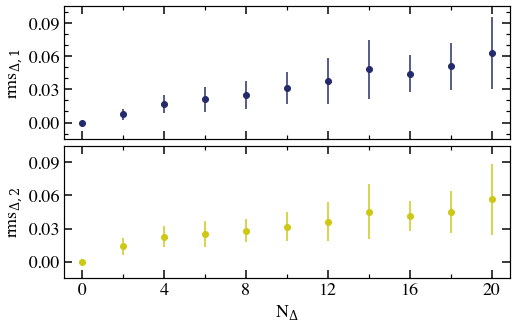}
\caption{Amount of distortion (rms) needed by the polynomials in order to fit the data to the model SEDs as a function of the number of distortion coefficients $\Delta_i$ used in the fit. }
\label{DistortionRMS}%
\end{figure}
We note a steady rise in the amount of distortion used with number of coefficients. A choice of $4 \leq N_{\Delta,i} \leq 10$ gives a balance between too little distortion and too much uncertainty in the amount of distortion needed, and rms$_{\Delta,i}$ is approximately constant in this range.
We cannot directly compare these rms differences with those between 1-D BT-Settl and 3-D {\sc stagger} models: the latter are defined over a narrower range ($0.2-20\,\mu$m) than the former (we use $0.1-30\,\mu$m). Restricting the wavelength range we use to match {\sc stagger} increases rms$_{\Delta,i}$ at least twofold as we lose useful constraints in the UV.

\begin{figure}
\centering
\includegraphics[width=0.475\textwidth]{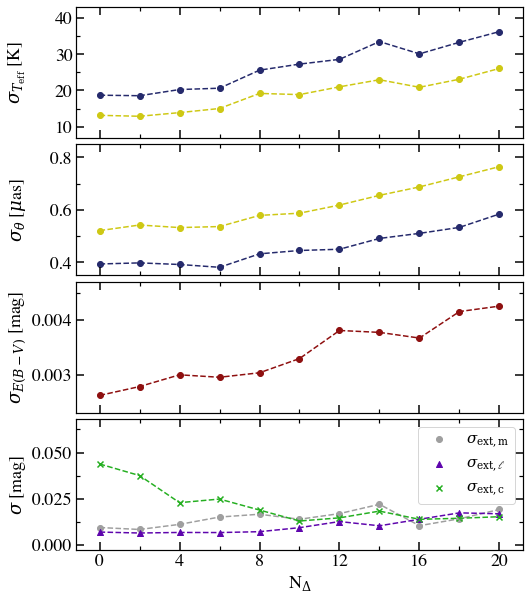}
\caption{Standard error on each fit parameter for each number of distortion coefficients $\Delta_i$. 
\textit{Top}: Errors on effective temperatures $\rm{T}_{\rm{eff},1}$ (blue) and $\rm{T}_{\rm{eff},2}$ (orange).
\textit{Second panel}: Errors on angular diameters $\theta_1$ (blue) and $\theta_2$ (orange).
\textit{Third panel}: Errors on interstellar reddening E(B$-$V).
\textit{Lower panel}: Errors on external error sources $\sigma_{\rm{ext,m}}$, $\sigma_{\rm{ext,}\ell}$ and $\sigma_{\rm{ext,c}}$.}
\label{DistortionParams}%
\end{figure}

Figure \ref{DistortionParams} shows the effect the number of distortion coefficients used has on the uncertainties in fit parameters.
The uncertainties in additional noise due to external sources of error, particularly $\sigma_{\rm ext,c}$, decrease with N$_{\Delta}$ while the uncertainties in physical parameters T$_{\rm eff}$, $\theta$ and E(B$-$V) increase. 
This could be explained by the hypothesis that for higher N$_{\Delta}$, the distortions begin to take advantage of flux ratios being unconstrained between filters and begin to move flux in and out of gaps between filters to improve the fit, which is not physically justified. The largest gap between filters we use is in the infrared (see Figure \ref{InputData}), but models are generally reliable in the IR (see IRFM) so we do not expect this to be a major issue in our final results. 
Our choice of N$_{\Delta}=10$ was made to balance the rising error in physical parameters with the high error in $\sigma_{\rm ext,c}$ at low orders.

\subsection{Different models}
We tested the effect of using models with different input parameters --- wavelength binning (runs A and B), temperature (run C) and metallicity (runs F and G). There is no significant effect in the result when changing the input SEDs. 
This indicates that our method of distortion is working and not very dependent on the input model.
In calculating our adopted values in run A, we used temperatures and a metallicity appropriate for AI Phe \citep{1988A&A...196..128A}, and smaller wavelength binning. There was no significant difference in the output between 20\,\AA\, and 50\,\AA\, bins, so we chose to run all other tests with the larger bin size in order to save computational costs.

\subsection{Constraining E(B--V) in the ultraviolet}
Constraints on interstellar reddening are vital in reducing the uncertainty on the derived values of effective temperature. 
In the case of AI Phe we were fortunate to be able to restrict the range of E(B$-$V) explored in our Bayesian analysis using the absence of the NaI doublet in HARPS spectra, and the depth of the 2175\,\AA\, absorption feature using the u220n and u220w ultraviolet fluxes and flux ratios. However, we looked at the effect of removing these constraints.
In run I of Table \ref{RunTable} we see a significant increase in the uncertainties on E(B$-$V) and T$_{\rm eff}$. If we remove the additional constraint provided by placing a narrow bandpass (u220n) in the UV, the uncertainties increase even further (run K).

Our results suggest that imposing a prior on E(B$-$V) is the strongest constraint on reddening, so including prior knowledge of interstellar reddening from e.g. spectroscopy or Stromgren photometry is a powerful way to improve results.
If this is not possible, having UV observations to pin down the shape of the UV end of the SED helps in determining the fit. By defining narrow bands in the IUE spectra, we were able to constrain the shape of the SED in the NUV more than if we used broad bands. These encompass more flux than narrow bands, but narrower bands are better at fixing the shape of the ultraviolet end of the SEDs.

\subsection{NIR flux ratios}
The T$_{\rm eff}$ values used in Section \ref{NIRFluxRatioSect} to derive the NIR flux ratio priors are calculated using the IFRM, so there may be some systematic offset between these values and the true T$_{\rm eff}$. Therefore, we need to check what the impact of an offset T$_{\rm eff}$-T$_{\rm IFRM}$ is on our results.
We found that our NIR flux ratio priors showed a very weak dependence on the values of T$_{\rm IFRM}$ used: introducing an offset of 100\,K changes the T$_{\rm eff}$ results by no more than $1-2$\,K (see run N in Table \ref{RunTable}). 
In general, applying NIR flux ratio priors to our data had little effect on our results for AI Phe.

\section{Conclusions}

For eclipsing binaries stars with well-defined eclipses it is possible to measure the radii of the two stars to much better than 1\% using high quality data --- 70 DEBs in DEBCat \citep{2015ASPC..496..164S} have masses and radii of both components measured to this accuracy. The end-of-mission accuracy of the parallaxes from the Gaia mission is expected to be at least 16\,$\mu$as for bright stars, i.e. better than 0.5\% for stars within 300\,pc. Therefore, precise and accurate angular diameters for many stars in eclipsing binaries are already available. There is potential to use these results to calculate precise and accurate fundamental effective temperatures for many stars, but this requires accurate measurements of the bolometric flux. The method that we present in this paper is a robust tool for deriving the bolometric flux for both stellar components in an DEB, provided there are enough photometric data.
We show the potential of this method with the well-characterised DEB, AI Phoenicis. The fundamental effective temperatures we obtained for this system are very precise: $T_{\rm eff,1}=6199\pm22$, $T_{\rm eff,2}=5094\pm16$. This is due to the high quality of the $R$ and $\varpi$ measurements, a strong upper limit on interstellar reddening, and constraints from ultraviolet photometry.
While the choice of input model SED has a small effect on the output effective temperatures, the tests we have done on the method show that uncertainties on the interstellar reddening have a large effect the uncertainty on the derived values of effective temperature. From the results in Table~\ref{RunTable}, we see that a constraint on E(B-V) of about 0.01\,mag is needed to reduce T$_{\rm eff}$ errors to less than 50\,K. Uncertain reddening will tend to bias the T$_{\rm eff}$ estimates as E(B-V) cannot be negative.

There are many bright eclipsing binaries of all types being discovered as a results of survey like WASP, KELT, K2, TESS, ASAS \citep{2018A&A...615A.135K,2017ApJ...844..134L,2018A&A...616A..38M,2019AJ....157..223L,2019A&A...622A.114H}, many of which have both Gaia parallaxes and a wealth of archival photometry. 
We conclude that the prospects for measuring accurate and precise effective temperatures for a large number of stars in eclipsing binaries are excellent.

\section*{Acknowledgements}

NM and PM are supported by the UK Science and Technology Facilities Council (STFC) grant numbers ST/M001040/1 and ST/S505444/1.
This research has made use of the SIMBAD database, operated at CDS,
Strasbourg, France. We would like to
thank the anonymous referee for their constructive and timely comments on the manuscript.

\section*{Data Availability}

No new data were generated or analysed in support of this research.



\bibliographystyle{mnras}
\bibliography{aiphe} 


\bsp	
\label{lastpage}
\end{document}